\documentclass[onecolumn, superscriptaddress]{revtex4-2}%
\UseRawInputEncoding
\usepackage{graphicx}
\usepackage{amssymb}
\usepackage{epstopdf}
\usepackage{graphicx}
\usepackage{amsmath}
\usepackage{amsfonts}
\usepackage{mathtools}
\usepackage{color}
\usepackage{changes}
\usepackage{lineno}

\begin{document}
\title{Nuclear dynamics of singlet exciton fission:\\ A direct observation in pentacene single crystals}
\author{H\'{e}l\`{e}ne Seiler}
\email[corresponding author: ]{seiler@fhi-berlin.mpg.de}
\affiliation{Fritz-Haber-Institut der Max-Planck-Gesellschaft, Berlin, 14195, Germany}
\author{Marcin Krynski}
\affiliation{Fritz-Haber-Institut der Max-Planck-Gesellschaft, Berlin, 14195, Germany}
\author{Daniela Zahn}
\affiliation{Fritz-Haber-Institut der Max-Planck-Gesellschaft, Berlin, 14195, Germany}
\author{Sebastian Hammer}
\email[corresponding author:]{ sebastian.hammer@physik.uni-wuerzburg.de}
\affiliation{Julius-Maximilians-Universit\"{a}t, Experimental Physics VI, 
Am Hubland, 97074 W\"{u}rzburg, Germany}
\author{Yoav William Windsor}
\affiliation{Fritz-Haber-Institut der Max-Planck-Gesellschaft, Berlin, 14195, Germany}
\author{Thomas Vasileiadis}
\affiliation{Fritz-Haber-Institut der Max-Planck-Gesellschaft, Berlin, 14195, Germany}
\author{Jens Pflaum}
\affiliation{Julius-Maximilians-Universit\"{a}t, Experimental Physics VI, 
Am Hubland, 97074 W\"{u}rzburg, Germany}
\author{Ralph Ernstorfer}
\affiliation{Fritz-Haber-Institut der Max-Planck-Gesellschaft, Berlin, 14195, Germany}
\author{Mariana Rossi}
\email[corresponding author:]{ mariana.rossi@mpsd.mpg.de} 
\affiliation{Fritz-Haber-Institut der Max-Planck-Gesellschaft, Berlin, 14195, Germany}
\affiliation{Max-Planck-Institut f\"{u}r Struktur und Dynamik der Materie, Hamburg, Germany}
\author{Heinrich Schwoerer}
\email[corresponding author:]{ heinrich.schwoerer@mpsd.mpg.de}
\affiliation{Max-Planck-Institut f\"{u}r Struktur und Dynamik der Materie, Hamburg, Germany}



\begin{abstract} 
Singlet exciton fission (SEF) is a key process in the development of efficient opto-electronic devices. An aspect that is rarely probed directly, and yet has a tremendous impact on SEF properties, is the nuclear structure and dynamics involved in this process. Here we directly observe the nuclear dynamics accompanying the SEF process in single crystal pentacene using femtosecond electron diffraction. The data reveal coherent atomic motions at 1 THz, incoherent motions, and an anisotropic lattice distortion representing the polaronic character of the triplet excitons. Combining molecular dynamics simulations, time-dependent density functional theory and experimental structure factor analysis, the coherent motions are identified as collective sliding motions of the pentacene molecules along their long axis. Such motions modify the excitonic coupling between adjacent molecules. Our findings reveal that long-range motions play a decisive part in the disintegration of the electronically correlated triplet pairs, and shed light on why SEF occurs on ultrafast timescales.
\end{abstract}

\pacs{}
\maketitle


Organic molecular semiconductors possess unique opto-electronic properties, combining the intrinsic optical characteristics of the individual molecules with the long range correlations enabled by intermolecular coupling. Among these properties, the ability of several organic semiconductors to undergo singlet exciton fission (SEF) has drawn tremendous fundamental and applied research interest over the past decades, summarized in several review articles \cite{Miyata2019, Smith2013, Smith2010, Casanova2018}. SEF is the process by which an electronically excited singlet exciton $S_1$ spontaneously splits into two triplet states $T_1 + T_1$. It may occur if the excess energy $\Delta E$, defined as the energy difference between the $S_1$ state and twice the lowest triplet state, is positive or if a small negative $\Delta E$ can be compensated by thermal energy. Owing to the ability to generate two electron-hole pairs per absorbed photon, the process bears high relevance for opto-electronic applications. \par
Since the discovery of SEF in the late sixties by measurements of the magnetic dependence of fluorescence \cite{Merrifield1969}, much emphasis has been placed on revealing the intermediate states and reaction pathways of the SEF process using time-resolved methods. The current understanding of the process involves three steps \cite{Miyata2019, Pensack2016}:
\begin{equation}
    S_1 \xrightleftharpoons{(1)} {^1(TT)} \xrightarrow{(2)} {^1(T..T)} \xrightarrow{(3)} T_1 + T_1.
    \label{eq:1}
\end{equation}
In this equation, step (1) describes the formation of the electronically and spin correlated triplet pair,  ${^1(TT)}$. The intermediate state ${^1(T..T)}$, formed during step (2), represents spatially separated triplet states, which are electronically independent but spin-correlated. Step (3) finally leads to two independent triplet excitons. Experiments on SEF dynamics have been almost exclusively based on photon excitation - photon probe schemes, such as transient absorption (TA) \cite{Wilson2011, Kolata2014, Birech2014}, two-dimensional spectroscopies \cite{Bakulin2015, Breen2017} or photoelectron spectroscopies \cite{Chan2011}, all being electronic probes. In equation \ref{eq:1}, step (2) confers most of their individual chemical and spectroscopic properties on the independent triplets, hence can be considered as the fission constituting process \cite{Miyata2019}. Due to its implications for applications, measuring its characteristic time constant is of importance. While step (2) has been previously studied in pentacene dimers using TA spectroscopy, timescale extraction has remained challenging due to spectrally overlapping transitions \cite{Pensack2016}. \par

SEF properties crucially depend on molecular structure and molecular packing in the crystal \cite{Kolata2014, Pensack2016, Janke:2020ff}. Fission yields and rates vary vastly with orientations and distances of neighbouring molecules, $\pi$-orbital correlations, $\Delta E$, and sample purity. Furthermore, recent ultrafast spectroscopy studies in combination with theory indicate that SEF is intrinsically linked to nuclear motion \cite{Miyata2017,Hart2018,Bakulin2015, Arag2015, Eggeman2013,Duan2020}. These studies call for an experimental probe that can access structural changes at the femtosecond timescale in a direct fashion.\par

\begin{figure*}[ht!]
    \centering
   \includegraphics[width = \linewidth]{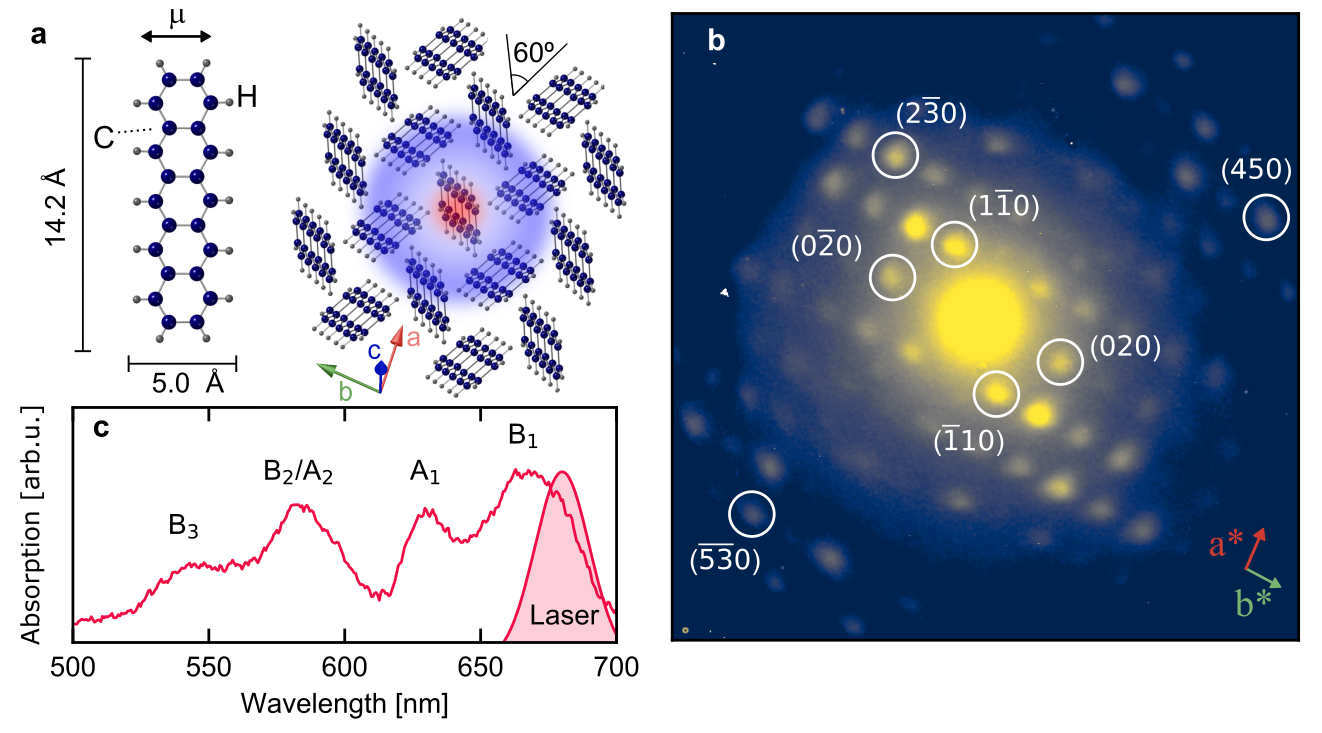}
    \caption{\textbf{Overview of the equilibrium structural and optical properties of the pentacene crystal}. \textbf{a} Illustrations of a single pentacene molecule (left) and of the crystal structure as viewed from a direction close to the $c$-axis (long axis). The red-blue halo depicts charge separation and $S_1$ exciton delocalization. The charge-transfer (CT) character of S$_1$ in crystalline pentacene has been established by various theoretical \cite{Sharifzadeh2013, Berkelbach2014,Cudazzo2012,Tiago2003,AmbroschDraxl2009, Cocchi2018} and experimental \cite{Hart2018} approaches. \textbf{b} Exemplary transmission electron diffraction pattern from a 40 nm thin, (001) pentacene single crystal slab. Intensity in log scale. \textbf{c} Unpolarized linear absorption spectrum of the pentacene sample. B$_1$, A$_1$ and B$_2$/A$_2$, B$_3$ are vibrational progressions of the higher and lower Davydov component at higher and lower energy, respectively, caused by a symmetric ring breathing mode of $\approx$ 170 meV, typical for polyacenes \cite{Hestand2015}. The shaded red peak area reflects the optical pump spectrum resonant with the S$_1\leftarrow$ S$_0$ transition.
}
    \label{fig:1}
\end{figure*}

In this work we directly reveal the structural dynamics accompanying the SEF process in single crystal pentacene using the method of femtosecond electron diffraction (FED), supported by real-time time-dependent density-functional theory (RT-TDDFT) and molecular dynamics simulations (MDS). Pentacene is the most widely studied SEF material and benefits from a wealth of previous theory and experimental works. Furthermore, the SEF process is exothermic at 300 K ($\Delta E \simeq$ 100 meV) and occurs with a yield close to 100\% \cite{Smith2013, Congreve2013}. 
By resonantly exciting the lowest singlet exciton, excess heating of the crystal is avoided. This allows isolation of the structural dynamics arising from the SEF process. Our experiments reveal coherent and incoherent contributions to the structural dynamics, as well as a long-lived ($\geq$ 1 ns), oriented structural distortion reflecting the polaronic character of the triplet excitons \cite{Hannewald2004}. 
The combination of RT-TDDFT, MDS and diffraction analysis enables us to identify the coherent motion as a delocalized, collective sliding motion along the pentacene's long axis. Further, the mechanism of coherent phonon generation is assigned to exciton-phonon and subsequent phonon-phonon coupling. Our findings imply that both coherent and incoherent motions participate in the SEF process in pentacene on the picosecond timescale by enabling the disintegration of ${^1(TT)}$. 

\section*{Results}
Pentacene (C$_{22}$H$_{14}$) single crystals were grown by sublimation, giving rise to a triclinic crystal structure (space group P-1, number 2) \cite{mattheus01, Siegrist2007}. In contrast to nanocrystalline thin films of varying polymorphs and defect concentrations, the sublimation method yields controllable growth of large single crystals with well-known crystal structure. The resulting herringbone structure, characteristic of polyacene crystals, is shown in Figure \ref{fig:1} (a). 
Details about sample growth and crystal structure are given in the Methods section. 
An exemplary transmission electron diffraction pattern is shown in Figure \ref{fig:1}(b), with the incident electron beam aligned along the crystal's $c$-axis. 
Pentacene being a pure hydrocarbon, the overall diffraction intensity is low compared to metal-organic compounds or organic-inorganic charge transfer complexes previously studied with FED \cite{Gao2013,Ishikawa2015,Smit2019,Jiang2017}. 
A diffuse scattering cloud is observed in the background, which we attribute to defects in the crystal and inelastic scattering. 

\subsection*{Femtosecond electron diffraction experiments}

\begin{figure*}[ht!]
    \centering
    \includegraphics[width = \linewidth]{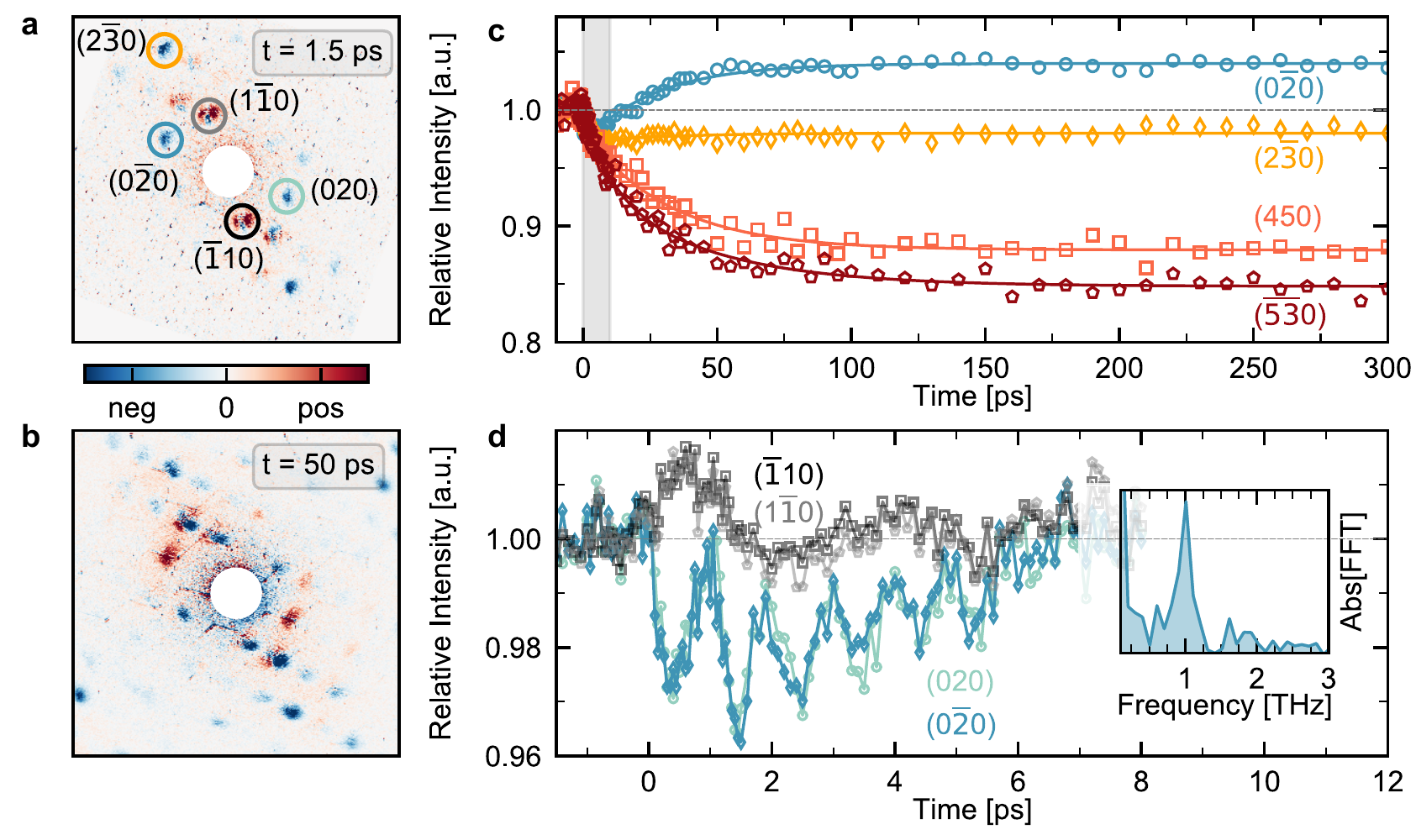}
    \caption{\textbf{Structural dynamics accompanying the singlet exciton fission process in pentacene.} \textbf{a.} Difference between the diffraction pattern at t = 1.5 ps after photoexcitation and prior to photoexcitation. \textbf{b.} Same as in \textbf{a}, but at t = 50 ps. \textbf{c.} Changes in intensity as a function of delay for selected Bragg reflections, indicated by circles in \textbf{a}. Fits are obtained from a global fitting procedure performed on the ensemble of reflections (Supplementary Figure 3). \textbf{d.} Zoom into the first picoseconds for the ($020$), ($0\overline{2}0$), ($110$) and ($1\overline{1}0$) reflections (shaded grey area in panel \textbf{c}). Inset shows the FFT spectrum of the oscillating residuals for ($0\overline{2}0$).}
    \label{fig:2}
\end{figure*}

For the FED measurements, we employ a 50 fs optical pump pulse with a central wavelength of 680 nm to resonantly excite the S$_1\leftarrow$ S$_0$ transition (B$_1$) and populate the lower Davidov component, see Figure \ref{fig:1} (c). 
The polarization of the pump pulse is tuned along the $a$-axis of the crystal, according to the direction of the transition dipole moment of S$_1\leftarrow$ S$_0$ (see Figure \ref{fig:1} (a) and Supplementary Figure 1) \cite{Hestand2015, Yamagata2011, Dressel2008}. 
We apply an incident fluence of 0.4 mJ/cm$^2$, yielding a calculated excitation density of 1 out of 30 molecules (Supplementary Note 1). 
The photo-excited singlet excitons subsequently undergo exothermic fission, and a femtosecond electron bunch probes the crystal lattice at a delay time $t$ after excitation, generating a diffraction pattern. 
This sequence is repeated for different delays. Further details about the FED instrument are available in the Methods section and elsewhere \cite{2015Wald}.\par  
An overview of the photo-induced changes in the diffraction pattern is presented in Figures \ref{fig:2}(a,b) in the form of intensity difference maps, obtained by subtracting the diffraction pattern before photo-excitation from the time-dependent patterns. 
Within our experimental resolution, no peak position shifts are observed. We thus treat size and shape of the unit cell as constant within the detected time window of 1 ns. 
However, we observe systematic intensity changes in various Bragg reflections, with certain intensities decreasing and others increasing following photoexcitation. 
Some orders, such as the ($020$) and ($1\overline{1}0$), also switch sign as the dynamics proceed. 
Since thermal heating leads to an intensity decrease of all Bragg reflections, the observed intensity changes cannot be caused by heating effects only. 
In addition, we estimate the average excess energy per molecule to only around 3 meV, which corresponds to heating under equilibrium conditions of the crystal by 1 K \cite{Fulem2008}. Therefore, beyond 50 ps, the observed intensity changes indicate a lattice distortion related to the electronic excitation (see Supplementary Note 2). The fact that the peak positions in reciprocal space do not change indicates that only a change in atomic positions within the unit cell occurs, which persists up to $> 1$ ns. \par

Figure \ref{fig:2}(c) shows time-dependent changes in intensity of few selected Bragg reflections. By performing a global fit to an ensemble of 16 time-dependent traces (see Supplementary Figure 3), we show that the structural dynamics can be decomposed into a sum of two exponential contributions, a fast picosecond component of 1.6 $\pm$ 0.2 ps and a slower component with a time constant of 28 $\pm$ 1 ps. These time constants unambiguously reveal the timescales of the incoherent lattice dynamics accompanying the SEF process in pentacene.

In order to assign the observed time constants to physical processes, we consider previously reported complementary studies. Electronic spectroscopies have largely focused on step (1) of Eq. \ref{eq:1}, which occurs within 100 fs in pentacene \cite{Chan2011,Wilson2011}. This step is below our instrument response function of $\simeq$ 250 fs, and we consider it instantaneous for our purposes. 
Step (3), the final breaking of the spin coherence between the triplet pair, was recently measured in time-resolved electron spin resonance experiments to be on the nanosecond timescale on tetracene, at least two orders of magnitude slower than the time constants observed here \cite{Weiss2016, Bayliss2019}. Hence, we tentatively attribute the fast incoherent structural dynamics to the disintegration of the electronically correlated triplet pair, ${^1(TT)} \xrightarrow{(2)} {^1(T..T)}$, and the slower one to its subsequent separation via coupling to delocalized vibrational modes and lattice disorder.\par

In addition to the incoherent structural dynamics, a closer look into the first ten picoseconds reveals pronounced oscillations at 1 THz of the amplitudes of several Bragg reflections on top of the incoherent dynamics. Oscillations are found to be strongest along the $b^*$-direction (indicated in Figure \ref{fig:1}(b)) as seen for the ($020$) and ($0\overline{2}0$) reflections in Figure \ref{fig:2}(d). While such oscillations are also present in several other diffraction orders (see Supplementary Figure 4), we do not observe them in the $a^*$-direction. 
A Fast Fourier Transform (FFT) of the oscillating parts of the Bragg reflections dynamics yields a peak at 1 $\pm$ 0.1 THz, shown in the inset of Figure \ref{fig:2}(d). The decay time of the oscillations is 3.9 ps $\pm$ 0.6 ps, and arises from energy dissipation and dephasing. To unravel the atomic motions behind the 1 THz oscillations in real space, we employ an approach that combines RT-TDDFT with MDS. We show how this approach, which allows for a direct comparison with the FED experiments \cite{Vasileiadis2019, Waldecker2017}, is well-suited given the complex vibrational structure of the pentacene crystal, featuring 108 phonon branches.

\subsection*{Theoretical modelling: nuclear dynamics}

\begin{figure*}[ht!]
    \centering
    \includegraphics[width = \linewidth]{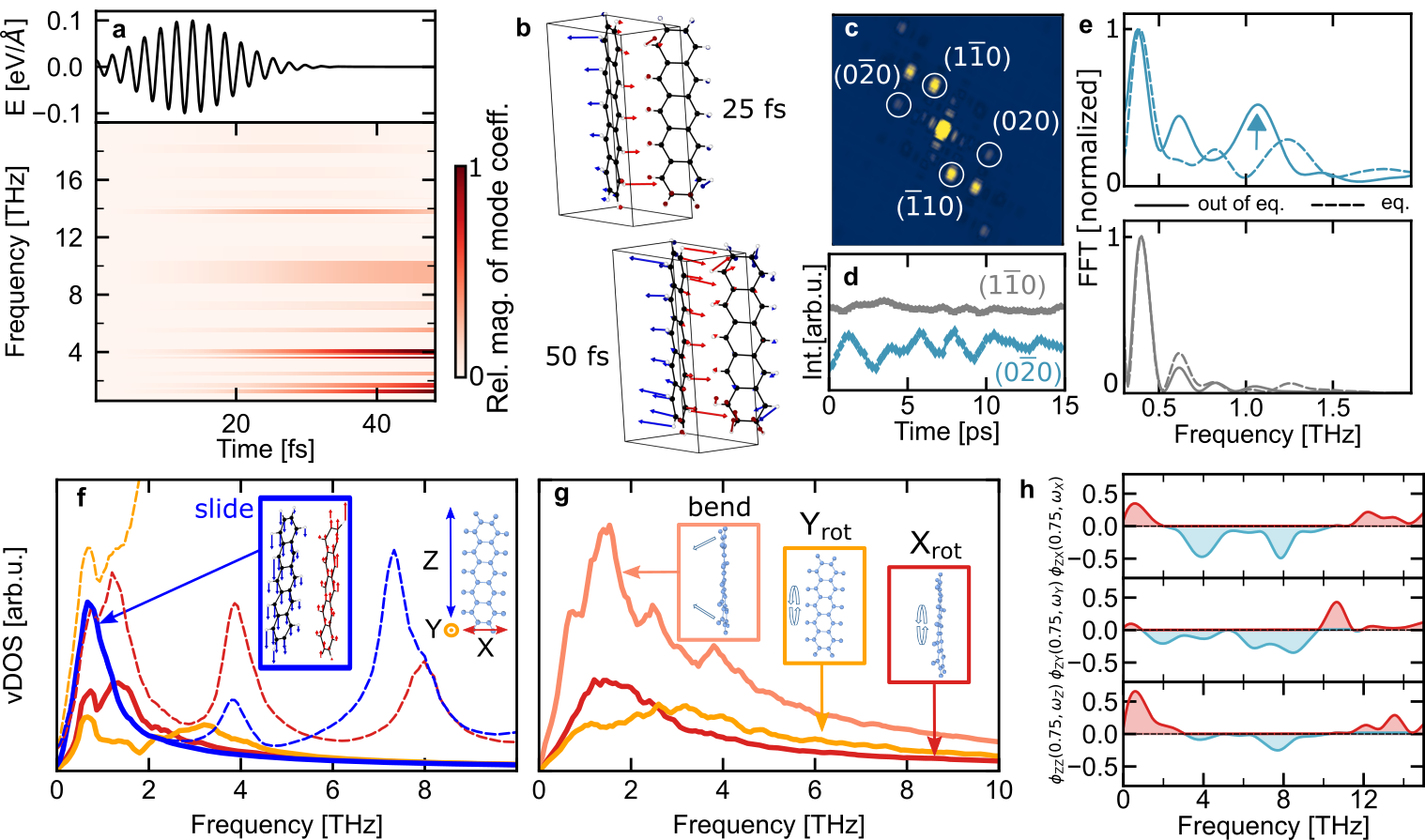}
    \caption
    {
    \textbf{RT-TDDFT combined with MD simulations yield insights into the collective motions accompanying singlet fission in real space.}
    \textbf{a} Projection of atomic displacements on the vibrational normal modes of the crystal, during Ehrenfest dynamics with a laser excitation as shown in the upper panel (RT-TDDFT). Color code indicates the relative magnitudes, at each time step, of each normal mode's coefficient. The harmonic frequencies of each mode are shown on the $y$ axis. \textbf{b} Real space atomic displacements of the crystal after 25 and 50 fs of Ehrenfest dynamics.
    \textbf{c} Simulated diffraction pattern from classical MDS.
    \textbf{d} Intensity fluctuations of simulated diffraction peaks from equilibrium MDS corresponding to peaks shown in Figure \ref{fig:2}(d). ($0\overline{2}0$): blue, ($1\overline{1}0$): grey.
    \textbf{e} FFT of the Bragg peak intensities shown in panel \textbf{d}, matching colors. Dashed: equilibrium MDS. Solid: out of equilibrium MDS. Data has been normalized. 
    \textbf{f} Partial equilibrium vDOS calculated for projections onto the $X$, $Y$ and $Z$ axes (inset). Dashed lines represent data obtained for carbon atoms only, solid lines are for the molecular CM. 
    \textbf{g} Partial vDOS patterns calculated for rotations of molecules around the $Y$ and $X$ axes and bending of the molecules (see text).
    \textbf{h} Selected slices (see text) from the 2D-vDOS auto- and cross-correlation patterns shown in Supplementary Figure 8, revealing phonon-phonon coupling between different directions.} 
    \label{fig:3}
\end{figure*}

To complement the FED data and investigate the complex structural dynamics of the pentacene crystal in real space, we perform two types of simulations, namely short \textit{ab initio} Ehrenfest dynamics including a laser excitation as modelled by RT-TDDFT, and long MDS in and out of equilibrium (using a hot-spot thermostat centered at 1 THz) with an empirical force field. Details about these simulations are given in the Methods section. 

The RT-TDDFT simulations are performed in a single pentacene crystal unit cell of the polymorph that matches the experimental diffraction pattern. These short time dynamics show that specific phonon modes, primarily in the region below $\simeq$ 15 THz, are activated following a laser excitation at 1.8 eV polarized along the $a$-axis, shown in Figure \ref{fig:1}(a). This is illustrated in Figure~\ref{fig:3}(a, b). Above 15 THz, we do not observe significant activation. The results in Figure \ref{fig:3}(a) indicates that two groups of normal modes are predominantly activated by the laser: around 1 THz and 4 THz. 
In Figure \ref{fig:3}(b) we show the atomic displacements observed after 25 fs and 50 fs of RT-TDDFT calculations. It is visible that the direction of these displacements is similar to the ones discussed in Ref.~\cite{Duan2020} for the singlet state, including a component in the direction of the mode at 5.4 THz. We find, however, that the atomic motion accompanying the singlet excitation in the anharmonic potential energy surface of the crystal is best described by a combination of several modes. All activated modes and a quantitative analysis of the coefficients at each snapshot is shown in Supplementary Figure 5.

Further insight on the character of the crystal dynamics in more realistic systems is obtained from the results of MDS in and out of equilibrium within a 3$\times$3$\times$2 pentacene crystal cell of the same polymorph. Out-of-equilibrium MDS were obtained from microcanonical dynamics following a thermalization 
by a hotspot thermostat \cite{Dettori2017} centered at
 1 THz.
A connection to the FED experiments can be obtained by simulating diffraction patterns for each MDS snapshot~\cite{Coleman2013, Vasileiadis2019}, as in Figure \ref{fig:3}(c). Figure \ref{fig:3}(d) displays simulated Bragg peak intensities as a function of time in equilibrium for the ($0\overline{2}0$) and ($1\overline{1}0$) reflections. The ($0\overline{2}0$) reflection fluctuates with a larger amplitude than the ($1\overline{1}0$) reflection. Panel (e) shows the FFT of these intensities, in equilibrium (dashed lines) and out of equilibrium (solid lines). A component around 1~THz of the ($0\overline{2}0$) Bragg reflection appears in the transient relaxation regime after vibrational excitation with the 1~THz-hotspot thermostat, in contrast to the ($1\overline{1}0$) reflection.

The characteristic motions around 1 THz deserve an in-depth analysis. Due to the  complexity of the collective motions observed in the MDS of the large crystal supercell (see movie S1), we present a decomposition of these motions into specific components, namely, rigid-body motions of the molecules in the crystal and low-frequency intramolecular motions.
This is achieved by calculating partial vibrational density of states (vDOS) from the Fourier transform of the velocity auto-correlation functions. 
In Figure \ref{fig:3}(f), we show the partial vDOS of simulations in thermal equilibrium (300 K) of the selected coordinates defined by the three orthogonal axes of each individual molecule, shown in the inset of Figure \ref{fig:3}(f): the direction parallel to the short edge of the molecule ($X$), the normal vector to the benzene rings of the molecule ($Y$) and the vector parallel to the long edge of the molecule ($Z$). Further details about the projection are provided in Supplementary Note 3. Specifically, Figure \ref{fig:3}(f) shows the projected vDOS of carbon atoms along $X$, $Y$ and $Z$ (dashed lines), as well as the vDOS of the center-of-mass (CM) along $X_{\textrm{CM}}$, $Y_{\textrm{CM}}$ and $Z_{\textrm{CM}}$ (solid lines).\par

These partial vDOS provide a clearer picture of the dynamical landscape. In particular, the $Z$ projection (dashed blue) is characterised by two sharp peaks around 1 THz and 8 THz. The 1 THz peak along this coordinate can be fully accounted by CM motions along $Z$ (solid blue line), as the dashed and solid blue lines overlap. 
This displacement of $Z_{\textrm{CM}}$ involves entire molecules sliding along each other in a highly collective, intermolecular motion which involves groups of molecules forming a wave along $Z$ (see Supplementary Figure 6 and movie S2).\par

The MDS results also show that a realistic picture of atomic motion of crystals in the THz region cannot be reduced to one single component. Below 2 THz  the vDOS exhibits components of CM motions not only along $Z$, as evidenced in Figure \ref{fig:3}(f). Further insights are obtained by tracking the CM motion of individual benzene rings in addition to the CM of entire molecules. This approach enables us to identify two rotations of the molecule around the $X$ and $Y$ directions, $X_{\textrm{rot}}$ and $Y_{\textrm{rot}}$, as well as an intramolecular bending motion of the molecule, $Z_{\textrm{bend}}$. The rotation around $X$ is related to the low-frequency mode that shows an activation in the RT-TDDFT simulations (see Figure \ref{fig:3}(b)). These motions, which all have a 1 THz component, are shown in Figure \ref{fig:3}(g). Together, they lead to collective intra and intermolecular motions of the molecules without altering the molecular CMs (see Supplementary Figure 7 and movie S3). 
In reality, the character of the vibrational modes that populate the 1 THz region at thermal equilibrium contain a combination of these motions, as expected in an anharmonic potential.\par

Finally, we find pronounced phonon-phonon coupling of the low THz region with higher-frequency phonon modes and between modes in different directions. 2D-vDOS correlation plots~\cite{Noda:00} of the motion along $X, Y$ and $Z$ coordinates were analysed, similarly to Ref.~\cite{Raimbault-PRM-2019}, where this method was also applied to molecular crystals. These plots are shown in Supplementary Figure 8, and selected cuts $\phi_{\alpha \beta}(\omega_\alpha,\omega_\beta)$, $\alpha,\beta=X, Y, Z$, are shown in Figure \ref{fig:3}(h). The red (blue) colour indicates positive (negative) correlation. These particular slices make it evident that: (i) The low-frequency (0.5-1.5 THz) sliding motions along $Z$, shown in inset of Figure \ref{fig:3}(f), couple to motions around 4 and predominantly 8 THz in the same direction (the 8 THz motions are vertical pulsing of the molecules); (ii) These same sliding motions along $Z$ strongly couple to motions along $X$ around $1$, $4$ and $8$ THz; (iii) They also couple to the motions along $Y$ in the regions between 2-4 and 6-8 THz. The motions along $Y$ and $X$, to which the sliding $Z$ motion couples, overlap with the ones populated upon the singlet excitation in the RT-TDDFT simulations. These results highlight the importance of phonon-phonon coupling in general for a realistic description of atomic motion in soft crystals.\par

\section*{Discussion}
\begin{figure*}[ht!]
    \centering
   \includegraphics[width = 0.8 \linewidth]{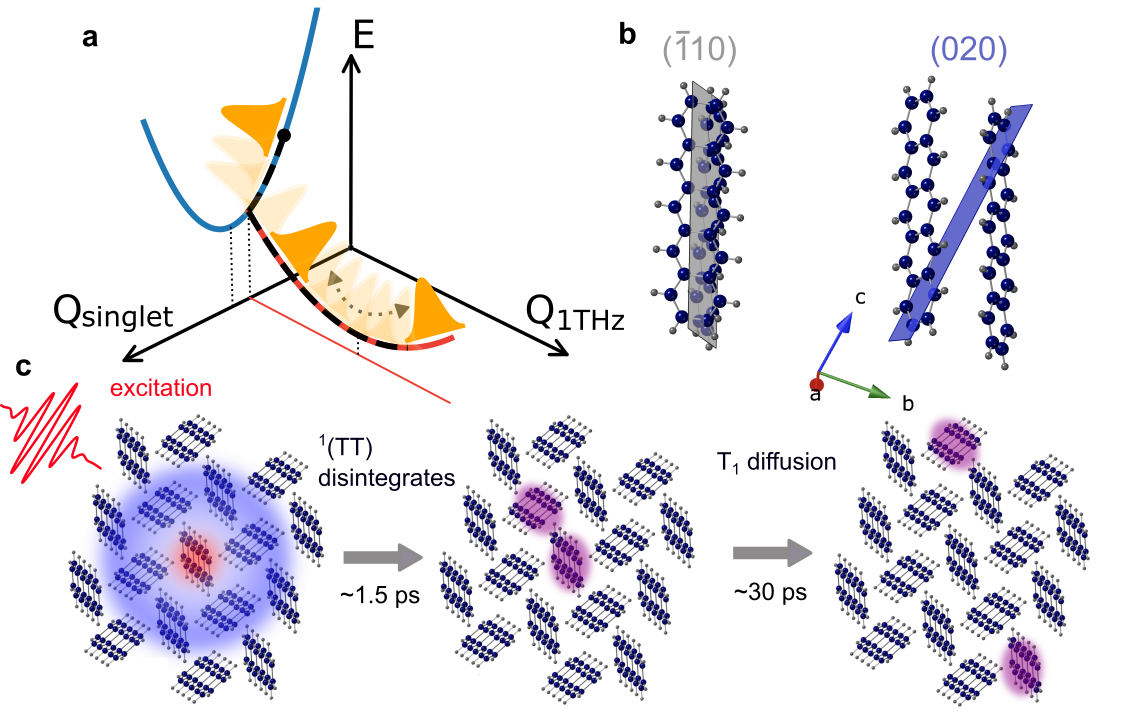}
    \caption{\textbf{Schematic illustrations of coherent and incoherent lattice dynamics of singlet fission in pentacene}. \textbf{a} Proposed mechanism of coherent phonon generation. The cartoon shows the excited potential energy surface $E$ and wavepacket motion (orange) on this surface. The laser excites inter and intramolecular modes (in particular around 4 THz) along the $Q_{\mathrm{singlet}}$ coordinate. These modes subsequently couple strongly to the 1 THz sliding motion via phonon-phonon coupling, initiating coherent wavepacket dynamics along the 1 THz coordinate Q$_{1\textrm{THz}}$. \textbf{b} The pentacene molecules cut through the (020) and the (\=110) planes differently. The 1 THz sliding motions appears much more pronounced in the (020) plane compared to the (\=110) plane.  \textbf{c} Possible real-space pictures associated with the observed incoherent structural dynamics. The purple halos represent triplet excitons.}
    \label{fig:4}
\end{figure*}

The combination of experiments and simulations yield new insights into the mechanisms of SEF in pentacene. By virtue of its strong CT character, photoexcitation of the $S_1$ exciton generates a local charge separation, with a partially positive central pentacene molecule, surrounded by four partially negative molecules \cite{Sharifzadeh2013}. The consequence of this charge separation is the instantaneous onset of a Coulomb force acting on the atoms of the crystal, which modifies their equilibrium position. Such a force explains the generation of coherent oscillations in the pentacene crystal via the displacive excitation mechanism (DECP) \cite{Zeiger1992}, illustrated in Figure \ref{fig:4}(a). Simulations reveal the microscopic mechanism of phonon generation. The RT-TDDFT simulations show that at short times, the laser excitation results in the population of inter and intramolecular modes in the crystal along the $a$ and $b$ axes, especially around 4 THz. The MDS reveal strong phonon-phonon coupling between these motions and the vertical sliding motion at 1 THz. In this picture, the coherent 1 THz motions thus arise from exciton-phonon and subsequent phonon-phonon coupling. \par

The consistency of the experimental signals with the sliding motion can be assessed using structure factor considerations. Modulations of a Bragg reflection intensity $I_{hkl}$ arise from modulation of the respective structure factor $F_{hkl} = \sum_j f_j \cdot {\rm e}^{{\rm i}\vec{G}_{hkl}\cdot\vec{r_j}}$, where $j$ is the $j^{\textrm{th}}$ atom of the unit cell, $\vec{r_j}$ its position, $f_j$ the atomic form factor, and $\vec{G}_{hkl}$ a Bragg scattering vector for Miller indices (hkl). Molecular motion perpendicular to a lattice plane (hkl) changes the respective scattering intensity $I_{hkl}$. In contrast, in-plane motions do not. 
We find that the 1 THz signal is observed in Bragg reflections for which the sliding motion has a perpendicular component to the lattice plane. No THz modulation is observed in Bragg reflections for which the sliding motion occurs within the lattice plane. Figure \ref{fig:4}(b) shows two examples: the $(020)$ plane, with a pronounced perpendicular component to the long axis of both inequivalent pentacene molecules and a pronounced 1 THz oscillation in the corresponding Bragg orders, and the $(\overline110)$ plane, parallel to the molecules' long axis, and a lack of 1 THz oscillation in the corresponding Bragg orders. All other investigated planes are shown in Supplementary Figure 4. The experimental structural dynamics are thus consistent with the long-axis sliding motion.\\
Previous studies on excitonic coupling in pentacene have demonstrated that thermal lattice fluctuations of such low-frequency modes, and in particular of the sliding motion \cite{Eggeman2013, Schweicher2019}, cause significant modulations of the electronic couplings between adjacent molecules, with direct influence on transport properties \cite{Arag2015, Eggeman2013, Schweicher2019}. These findings also apply to non-equilibrium situations, such as in the SEF process. Therefore, we expect that the 1 THz sliding motion, as well as their coupled low-frequency motions, play an important role in the disintegration of the electronically correlated triplet pair ${^1(TT)}$ on the picosecond timescale. \par

In addition to the coherent structural dynamics, FED measurements cleanly isolated the incoherent contributions of lattice dynamics to the SEF process. The $1.6$ ps incoherent dynamics, assigned to the fission constituting process, act in parallel with the coherent dynamics and also contribute to ${^1(TT)}$ disintegration. The second incoherent time scale is attributed to triplet diffusion, reaching an equilibrium spatial distribution within $\simeq$ 30 ps. This timescale, much faster than spin dephasing \cite{Weiss2016, Bayliss2019}, is indeed on the same order of magnitude as triplet exciton diffusion timescale in the $ab$-plane of anthracene crystals \cite{Ern1969, Avakian1968}. Finally, we discuss the potential origin of the structural distortion shown in Figure \ref{fig:2}(b). The intensity rise observed in the $(020)$ and ($0\overline{2}0$) reflections over that timescale, while no structure factor changes are observed in the (h00) reflections, indicate that the structural distortion effectively happens along the $b$-axis of the crystal. Because this axis is the mean direction between non-equivalent pentacene neighbours ($b+a$/2 and $b-a$/2), and these non-equivalent pairs have the highest $\pi$-orbital overlap, one would expect attraction or repulsion caused by the long living triplet state interacting with its environment to induce exciton-polarons primarily in the $b$ direction.\par

In summary, we have employed FED to directly probe the photo-induced lattice dynamics in a prototypical SEF material. Our findings imply that both coherent and incoherent motions participate in the SEF process in pentacene on the picosecond timescale. The dominant atomic motions as well as the mechanism for coherent phonon generation have been identified by simulations and were found consistent with the experimental data. This study shows that including long-range intermolecular coupling is essential for an accurate description of structural dynamics of SEF in pentacene, and considering motions within the unit cell only is not sufficient. We expect this finding to be relevant more generally for molecular single crystals and polycrystalline thin films, as well as soft crystals like lead-halide perovskites. 

\section*{Methods}
\subsection{Pentacene crystal growth}
Pentacene single crystals (C$_{22}$H$_{14}$) were grown via horizontal physical vapor deposition \cite{PVD1, PVD2} using about 50 mg of twofold gradient sublimation purified pentacene. The starting material is placed in a horizontal furnace with a well defined temperature gradient. The material is sublimed at 290$^\circ$C and transported by a continuous 30 sccm N$_2$ (6N purity) inert gas flow along the temperature gradient, leading to condensation of thin plate-like single crystals from supersaturated vapor in the colder zone of the furnace. The as-grown crystals were slowly cooled down over 12 hours to minimize thermal stress. By use of a microtome, crystals were subsequently cut parallel to their {\it (ab)}-facets into platelets of 100's of $\mu$m lateral size and 30 to 80 nm thickness to be used in FED experiments (see Supplementary Figure 9). From out-of-plane x-ray diffraction (XRD) measurements the common pentacene bulk crystal phase, characterized by the (001) lattice distance of 14.1 Å, is identified \cite{Hammer2020,mattheus01, Siegrist2007}. Comparison of the experimental electron diffraction pattern of the cut crystal platelets shown in Figure \ref{fig:1}(b) with simulated patterns allows us to determine the single crystal structure to be in the so-called high temperature Siegrist phase, which is know to be metastable at room temperature \cite{Siegrist2007}. As the diffraction pattern obtained by FED can be unambiguously assigned to this high temperature phase for various samples, we conclude that the microtrome cutting induces a phase transition to the high temperature phase in the platelet without changing the crystalline integrity of the sample as clearly shown by the high quality FED pattern. 

\subsection{Femtosecond Electron Diffraction experiments}
The output of a Ti:Sapph amplifier (Coherent, Astrella, 4 kHz, 6W, 50 fs) is split into a pump and a probe path. The pump is derived from a commercial optical parametric amplifier (TOPAS, Light Conversion). For the probe, a home-built NOPA is employed followed by a prism-compressor setup for dispersion compensation. The roughly 400 nJ NOPA pulses are focused onto a gold photocathode for the generation of femtosecond bunches of photoelectrons. The photoemitted electrons are accelerated towards the anode at 70 keV. More details about the electron gun design are provided in Ref. \cite{2015Wald}. We minimize the probe electrons' propagation path to the sample, so as to minimize temporal broadening of the electron pulse due to space-charge effects. The electron bunch diffracts off the sample and a magnetic lens is employed to focus the electrons onto a detector (TVIPS, F416). In this experiment, the temporal resolution is $\simeq$ 250 fs. 
To analyze the Bragg reflection intensities quantitatively, each peak in the diffraction pattern is fitted with a 2D-Pseudo Voigt function and a tilted background surface (see Supplementary Figure 2).

\subsection{Theoretical modelling}

The time-dependent density functional theory (TDDFT) calculations performed in this project were carried out with the Octopus code \cite{doi:10.1063/1.5142502, https://doi.org/10.1002/pssb.200642067} with the Perdew-Burke-Ernzerhof \cite{PhysRevLett.77.3865} exchange and correlation functionals and D3 \cite{doi:10.1063/1.3382344} corrections for the van der Waals forces, and including spin polarization. In the first step, the pentacene crystal structure (primitive cell, 2 molecules) was optimised using the FIRE algorithm \cite{PhysRevLett.97.170201}, as implemented in Octopus, where a 0.16
~\AA~grid spacing was used. The optimization was stopped when no forces were higher than 0.01 eV/{\AA}.
Next, the laser-induced dynamics was simulated by applying a time-dependent electric field on the ground-state structure with 1.8 eV carrier frequency, maximum amplitude of 0.1 eV/{\AA}, a Gaussian-pulse shape with $\tau_0 = 6.6$ fs, resulting in a pulse duration of approximately 25 fs. Ehrenfest dynamics were propagated with a time step of 0.001 fs and the total simulation time was of 50 fs.

Molecular dynamics trajectories were obtained using the AIREBO interatomic force-field \cite{doi:10.1063/1.481208} within the LAMMPS code \cite{lammps}. A 3$\times$3$\times$2 simulation cell (1152 atoms) with periodic boundary conditions is employed in order to ensure a good description of low-frequency phonon modes. By performing a comparsion of harmonic phonons, we have found that the frequencies of equivalent phonons modes with the force-field and DFT can differ by up to 1 THz, likely due to the approximate nature of the empirical potential.
All MDS calculations were carried out at 300 K, with 0.5 fs steps.
Equilibrium MDS trajectories were obtained in a two-step procedure. First, a canonical ensemble simulation using the stochastic velocity rescaling thermostat~\cite{Bussi_2007} was performed. 50 ps of continuous trajectory was obtained excluding 10 ps of initial thermalisation.
Next, six uncorrelated frames from thermalised NVT trajectory were extracted and used as initial configurations (positions and velocities) for consecutive NVE microcanonical simulations. The latter was performed for 15 ps.
Out of equilibrium MDS simulations were performed using a hot-spot, non-equilibrium generalized Langevin equation thermostat \cite{Dettori2017} to selectively excite to 400 K a narrow range of vibrational modes in the area of 1 THz, keeping the rest of the system at a constant temperature of 300 K.
50 ps of continuous trajectory was obtained excluding 10 ps of initial thermalisation.
Next, six frames from hot-spot simulation trajectory were extracted and used as initial configurations (positions and velocities) for consecutive NVE microcanonical simulations. These simulations were performed for 15 ps and this data was analyzed.

\section*{Acknowledgements}
This work was funded by the Max Planck Society, the European Research Council (ERC) under the European Union’s Horizon 2020 research and innovation program (Grant Agreement Number ERC-2015-CoG-682843), and by the Deutsche Forschungsgemeinschaft (DFG) - Projektnummer 182087777 - SFB 951 (B17 and A13). H.S. ~acknowledges support by the Swiss National Science Foundation under Grant No.~P2SKP2\textunderscore184100. Y.W.W acknowledges funding from the DFG within the Emmy Noether program under Grant No. RE 3977/1. M.K. acknowledges funding from the BigMAX Max Planck research network on big-data-driven material science. S.H. and J.P. are grateful to the DFG for support within project PF385/12 as well as to the Bavarian State Ministry for Science and the Arts for funding within the collaborative research network "Solar Technologies go Hybrid" (SolTech). We thank Shawn Coleman for helpful discussions.
We acknowledge useful discussions and help with the Octopus code from Heiko Appel, Nicolas Tancogne-Dejean and Matheus Jacobs.

\section*{Author Contributions}
H.S., R.E, and H.Sch. conceived the study; H.S., H.Sch. conducted the experiments with contributions from D.Z.; Pentacene crystal growth and characterization was performed by S.H. and J.P.; Experimental data were analysed by H.S. and H.Sch. with contributions from R.E., D.Z, Y.W.W. and T.V.; M.R. and M.K. designed the theoretical modelling and analysed the results from simulations; M.K. performed all theoretical calculations and post-processing of simulation data; H.S., H.Sch., M.K. and M.R. wrote the manuscript. All authors discussed the results and commented on the manuscript. 

\section*{Competing Interests}
The authors declare no competing interests.
\bibliographystyle{naturemag}
\bibliography{bibliography}

\end{document}


\title{Supplementary Materials \\Atomic motions accompanying singlet fission in pentacene single crystals}
\title{Supplementary Materials \\Nuclear dynamics of singlet exciton fission:\\ A direct observation in pentacene single crystals}
\author{H\'{e}l\`{e}ne Seiler}
\email[corresponding author: ]{seiler@fhi-berlin.mpg.de}
\affiliation{Fritz-Haber-Institut der Max-Planck-Gesellschaft, Berlin, 14195, Germany}
\author{Marcin Krynski}
\affiliation{Fritz-Haber-Institut der Max-Planck-Gesellschaft, Berlin, 14195, Germany}
\author{Daniela Zahn}
\affiliation{Fritz-Haber-Institut der Max-Planck-Gesellschaft, Berlin, 14195, Germany}
\author{Sebastian Hammer}
\email[corresponding author: ]{ sebastian.hammer@physik.uni-wuerzburg.de}
\affiliation{Julius-Maximilians-Universit\"{a}t, Experimental Physics VI, 
Am Hubland, 97074 W\"{u}rzburg, Germany}
\author{Yoav William Windsor}
\affiliation{Fritz-Haber-Institut der Max-Planck-Gesellschaft, Berlin, 14195, Germany}
\author{Thomas Vasileiadis}
\affiliation{Fritz-Haber-Institut der Max-Planck-Gesellschaft, Berlin, 14195, Germany}
\author{Jens Pflaum}
\affiliation{Julius-Maximilians-Universit\"{a}t, Experimental Physics VI, 
Am Hubland, 97074 W\"{u}rzburg, Germany}
\author{Ralph Ernstorfer}
\affiliation{Fritz-Haber-Institut der Max-Planck-Gesellschaft, Berlin, 14195, Germany}
\author{Mariana Rossi}
\email[corresponding author: ]{ mariana.rossi@mpsd.mpg.de} 
\affiliation{Fritz-Haber-Institut der Max-Planck-Gesellschaft, Berlin, 14195, Germany}
\affiliation{Max-Planck-Institut f\"{u}r Struktur und Dynamik der Materie, Hamburg, Germany}
\author{Heinrich Schwoerer}
\email[corresponding author: ]{  heinrich.schwoerer@mpsd.mpg.de}
\affiliation{Max-Planck-Institut f\"{u}r Struktur und Dynamik der Materie, Hamburg, Germany}

\pacs{}
\maketitle

\newpage
\section{Supplementary note 1: estimation of excitation density}

The aim is to determine the fraction $\eta$ of pentacene molecules which are excited by the incident laser pulse. We start by considering the photon energy of 1.82 eV (680 nm):
\begin{equation}
E_{\rm ph} = \frac{\rm hc}{\lambda} = \frac{2\cdot 10^{-25}\,\rm Jm}{680\cdot 10^{-9}\,\rm m} = 2.9\cdot 10^{-19} \rm J  
\end{equation}
The incident number of photons per area is:
\begin{equation}
N_{\rm ph,in} = \frac{F_{\rm i}}{E_{\rm ph}} =  \frac{0.4\cdot 10^{-3} \rm J/cm^{-2}}{2.9\cdot 10^{-19} \rm J} \approx  1.4  \cdot 10^{15} \rm cm^{-2},
\end{equation}
Where $F_i \simeq$ 0.4 mJ/cm$^{-2}$ is the incident fluence on the sample. Next we calculate the absorbed number of photons per volume. The fraction of absorbed photons is $\alpha = 1 - 10^{-\epsilon(\lambda)\cdot d}$, with  the absorption coefficient in unit per length $\epsilon$\ and the thickness $d$.
We take $\epsilon(\lambda)\cdot d$ from linear absorption measurements. For example, a 50 nm thick crystal at the low Davidov component at 680 nm yields $\epsilon(\lambda)\cdot d = 0.20$.  Hence the number of absorbed photons per unit volume is: 
\begin{equation}
N_{\rm ph,abs} = \frac{\alpha(680 \rm nm)\cdot N_{\rm ph,in}}{d} = \frac{0.37\cdot 1.4 \cdot 10^{15} \rm cm^{-2}}{50 \,\rm nm} \approx 1 \cdot 10^{20} \,\rm cm^{-3},
\end{equation}
Taking into account the unit cell volume of $V_{\rm uc} = $ 705 \AA$^3$, and a factor 0.5 because each unit cell contains 2 pentacene molecules, we obtain an excitation density of:
\begin{equation}
\eta = N_{\rm ph,abs}\cdot V_{\rm uc} = 1.0 \cdot 10^{20} \,\rm cm^{-3} \times 350 \cdot 10^{-24}\,\rm cm^3 \approx 0.036
\end{equation}
Roughly 3\% of the molecules are excited with our incident fluence and pump central wavelength 680 nm. Even though only a thirtieth of the molecules are excited, these conditions yield a high density of excitations given the delocalized nature of the first singlet exciton state. According to Sharifzadeh et al., this would mean a distance of about 3 molecules between excitations \cite{Sharifzadeh2013}. 

\newpage
\section{Supplementary Note 2: Temperature rise and Debye-Waller effect}
We first estimate the temperature increase resulting from laser excitation, using the excitation density $(\eta = 1/30)$, the specific heat $C$ of pentacene \cite{Fulem2008}, and assuming a fission yield of 100$\%$ (i.e. neglecting singlet photoluminescence, which is extremely weak in crystalline pentacene \cite{Bowen_1949}). To estimate the energy in the system after photoexcitation, we consider the following: the maximum amount of energy that can end up in the lattice is the full photon energy (1.82 eV, corresponding to a 680 nm photon), which would result in a lattice temperature rise of $\Delta T \simeq$ 20 K (equation (\ref{deltat})). However, ultrafast non-radiative decay is unlikely for triplets in pure crystalline hydrocarbons \cite{Siebrand1966, Robinson1962}. Indeed the dominant decay channel is  phosphorescence with a timescale much longer than the observation window here \cite{Siebrand1967}. Therefore it is reasonable to assume that the photon energy is mostly conserved in the triplet excitons. Within these assumptions, the only energy remaining to heat the sample is the energy  difference between $E(S_1$) and $2\cdot E(T_1)$. For the lower Davidov component in pentacene, this quantity is $\simeq$ 100 meV. Hence we have:
\begin{eqnarray}
\label{deltat}
\Delta T & = & \frac{E_{\rm abs}}{C}\, , \quad C = 300\, \frac{\rm J}{\rm K\cdot mol}  \\[1ex]
& & E_{\rm abs} = \frac{3\; {\rm meV}}{\rm molecule} = 18\cdot 10^{23} \frac{\rm meV}{\rm mol} \approx 300 \frac{\rm J}{\rm mol}\\[2ex]
\Delta T & = & \frac{300{\rm J\cdot mol\cdot K}}{300{\rm J\cdot mol}} \approx 1\; \rm K.
\end{eqnarray}

We estimate the changes of the Debye-Waller (DW) factor for the thermal equilibrium temperature increase $\Delta T$  following photo-excitation of the sample. Here we are looking for an order of magnitude of the DW effect, as opposed to a quantitative description which goes beyond the scope and need of this work. We use a Debye model with a Debye temperature of 18 meV for pentacene \cite{Hatch2010}. In this estimation we use the atomic mass of carbon, since their contribution to the diffraction intensity is much larger compared to the hydrogen atoms. We first compute the change in mean-squared atomic displacement (MSD) for $\Delta T \approx 1 K$. We obtain:
\begin{equation}
\Delta \langle u \rangle ^2 = \langle u \rangle^2( \mathrm{T} = 301 \mathrm{ K}) - \langle u \rangle^2 (\mathrm{T} = 300 \mathrm{ K}) \approx 5 \cdot 10^{-24} \mathrm{\AA}^2
\end{equation}

Next we establish how much this MSD change impacts the relative Bragg peak intensities. The intensity of a Bragg reflection is given by the square of the structure factor $I_{hkl} = |F_{hkl}|^2$. For isotropic vibrations, the Debye-Waller corrected structure factor is $F_{hkl}= \sum_j f_j \cdot {\rm e}^{{\rm i}\vec{G}_{hkl}\cdot\vec{r_j}}{\rm e}^{-\frac{1}{3}\langle u \rangle^2 \cdot |\vec{G}_{hkl}|^2}$, where $j$ is the $j^{\textrm{th}}$ atom in the unit cell, $f_j$ the atomic form factor, $\vec{r_j}$ the position of atom $j$ in the unit cell and $\vec{G}_{hkl}$ the Bragg scattering vector for Miller indices (hkl). For the (020) reflection, for example, we find that $I_{020}$(T = 301 K)/$I_{020}$(T = 300 K) = $e^{-\frac{1}{3}\Delta \langle u \rangle^2 \cdot |\vec{G}_{020}|^2} \approx $ 0.999, with $|\vec{G}_{020}| \approx 1.6 \cdot 10^{10}\,{\rm m}^{-1}$. This means that heating effects are expected to have a marginal though not zero influence on the transient intensities.

\newpage
\section{Supplementary note 3: projection used in the partial vDOS calculations}
We provide details of the decomposition along the $X$, $Y$, $Z$ shown in Figure 3(f) of the main text. To extract the partial vDOS from the MD simulations along these axes, we have applied a projection scheme in which each molecule is treated separately and is transformed by translation ($\bm{T}$) and rotation ($\bm{M}_1$, $\bm{M}_2$) operations. After these transformations, each molecule lies in the $x-z$ plane of Cartesian coordinates, with the long-edge of the molecule parallel to the $z$ axis. The $X$, $Y$, and $Z$ components are then easily defined according to this orientation of the molecules and the rotation and translation matrices are kept. In detail, the procedure was achieved by performing the following steps for each molecule in the simulation cell:

\begin{enumerate}
  \item The average atomic positions of all atoms of each pentacene molecule ($\bar{\bm{x}}$, $\bar{\bm{y}}$, $\bar{\bm{z}}$) over the entire MD simulation time were calculated. These positions were used to extract translation matrices.
  \item Specifically, the translation matrix $\bm{T}$ was obtained from the translation operation of the molecular center-of-mass (CM) ($\bar{x}_{CM}$, $\bar{y}_{CM}$, $\bar{z}_{CM}$) to the origin. This matrix was then used to shift the ($\bar{\bm{x}}$, $\bar{\bm{y}}$, $\bar{\bm{z}}$).
  \item The translated ($\bar{\bm{x}}^T$, $\bar{\bm{y}}^T$, $\bar{\bm{z}}^T$) were used to perform a  singular value decomposition (SVD). The angle between primary vector $\bm{v}_1$ and the Cartesian direction $z$ was obtained and used to calculate rotation matrix $\bm{M}_1$. Next, the atomic positions were rotated with the rotation matrix $\bm{M}_1$ resulting in new atomic positions ($\bar{\bm{x}}^{M_1T}$, $\bar{\bm{y}}^{M_1T}$, $\bar{\bm{z}}^{M_1T}$) such that the long edge of the molecule aligns with the $z$ axis.
  \item Another SVD was performed on ($\bar{\bm{x}}^{M_1T}$, $\bar{\bm{y}}^{M_1T}$, $\bar{\bm{z}}^{M_1T}$). This time the secondary singular vector $\bm{v}_2$ was extracted, as well as the angle between $\bm{v}_2$ and the $x$ axis. The obtained angle was used to calculate rotation matrix $\bm{M}_2$ and to further rotate the atomic positions, so that the short edge of resulting molecule with positions  ($\bar{\bm{x}}^{M_1M_2T}$, $\bar{\bm{y}}^{M_1M_2T}$, $\bar{\bm{z}}^{M_1M_2T}$) aligns with the $x$ axis.
  \item Finally, the obtained $\bm{T}$, $\bm{M}_1$ and $\bm{M}_2$ matrices were used to transform the atomic positions of each simulation snapshot of a given molecule. 
\end{enumerate}

Based on the atomic positions transformed with the $\bm{T}$, $\bm{M}_1$ and $\bm{M}_2$ matrices, the velocities were calculated as the time derivative of the transformed positions with a time interval of 1 fs, and subsequently used to calculate the projected vDOS patterns shown in the main text.

\newpage
\section{Supplementary Figure 1}

\begin{figure}[ht!]
    \renewcommand{\figurename}{Supplementary Figure}
    \centering
    \includegraphics{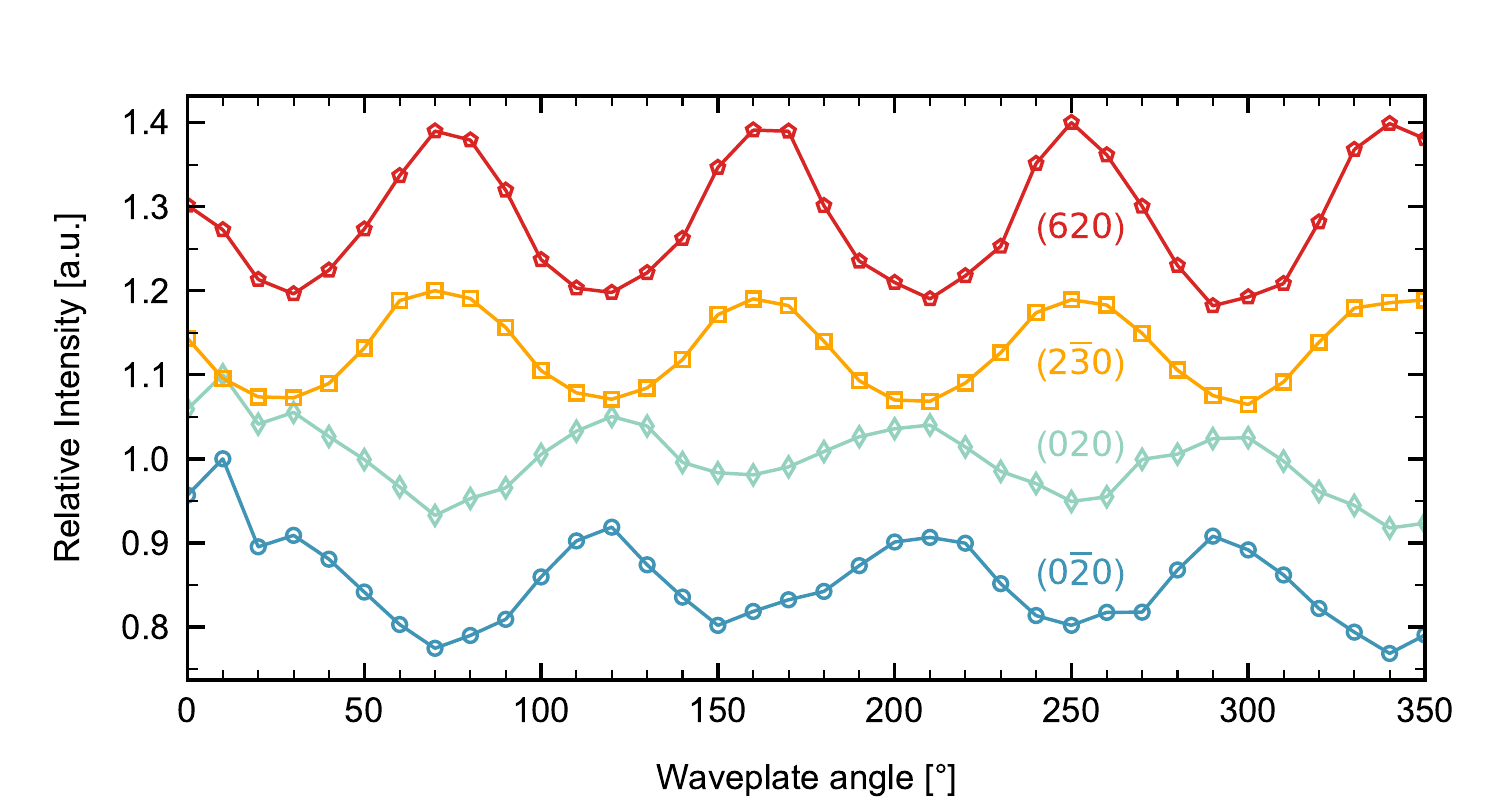}
    \caption{\textbf{Tuning the laser light polarization with respect to the sample.} The intensity of Bragg reflections are monitored as a function of waveplate angle in the pump path, at a pump central wavelength of 680 nm and a constant pump-probe delay of 50 ps. Graphs are offset for clarity. Intensity modulations of the Bragg reflections are used as a proxy for the energy absorbed by the sample, a quantity which hugely varies as a function of light polarization in pentacene single crystals. Indeed polarization-dependent linear absorption measurements have revealed a much enhanced absorption along the $a$-axis of the crystal \cite{Cocchi2018}. Here, absorption along $a$ corresponds to a waveplate angle of 120$^{\circ}$, with ($620$) reaching a minimum and ($0\overline{2}0$) reaching a maximum (indicated by the dashed line).}
    \label{fig:s1}
\end{figure}
\newpage

\section{Supplementary Figure 2}
Each Bragg reflection in the diffraction pattern is fitted with a 2D Pseudo-Voigt function with a tilted background surface:
\begin{equation*}
z(x,y) = a\cdot \Big [(1- \eta)\cdot e^{-\textrm{log}(2)\cdot \big [((x-x_0)/2\sigma_x)^2 + ((y-y_0)/2\sigma_y)^2\big ]} + \eta \cdot \frac{1}{[1 + ((x-x_0)/2\sigma_x)^2 + ((y-y_0)/2\sigma_y)^2\big ] } \Big ] + b_1 \cdot x + b_2 \cdot y + b_3
\end{equation*}
In this expression, $a$ is the peak amplitude, $\eta$ is the mixing ratio between Lorentzian and Gaussian, $x_0$ and $y_0$ are the peak center positions, $\sigma_x$ ($\sigma_y$) is the width in the $x$ ($y$) direction, and $b_1$-$b_3$ are the tilted background plane parameters. An example peak fitting result obtained with $z(x,y)$ is displayed in Supplementary Figure \ref{fig:fit} for the ($\overline{1}20$) peak. The fitting procedure was performed on all the analyzed Bragg reflections at all delays, and the value of the parameter $a$ was used to report on the relative intensity changes in Figure (2) of the main manuscript.

\begin{figure}[ht!]
    \renewcommand{\figurename}{Supplementary Figure}
    \centering
    \includegraphics{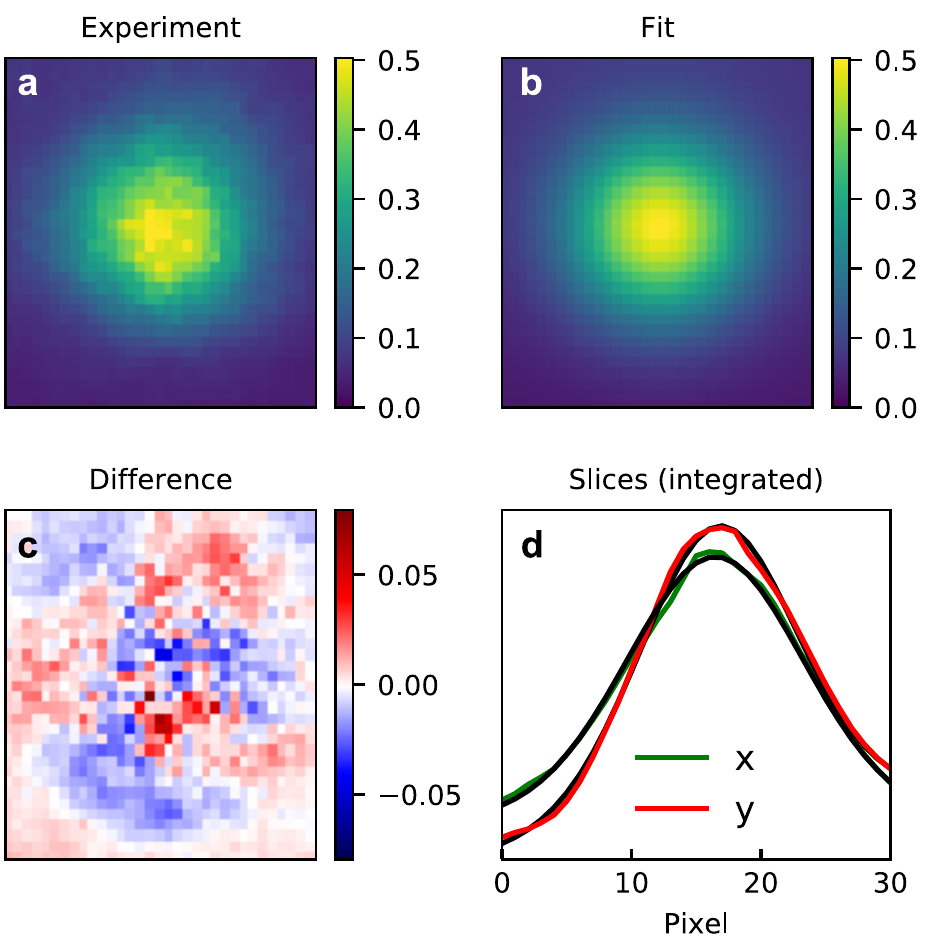}
    \caption{\textbf{Example fit of a 2D Pseudo-Voigt with a tilted background surface on the ($\overline{1}20$) Bragg reflection.} \textbf{a.} Experimental data. A small region of interest around the ($\overline{1}20$) peak is selected. \textbf{b.} Fit obtained by the function $z(x,y)$ \textbf{c.} Difference between the data and the fit. \textbf{d.} Integrated experimental signal along the x (green) and y (red) dimensions of the image and corresponding fit (black).}
    \label{fig:fit}
\end{figure}

\newpage

\section{Supplementary Figure 3}
\begin{figure}[ht!]
    \renewcommand{\figurename}{Supplementary Figure}
    \centering
    \includegraphics[width = 0.9 \linewidth]{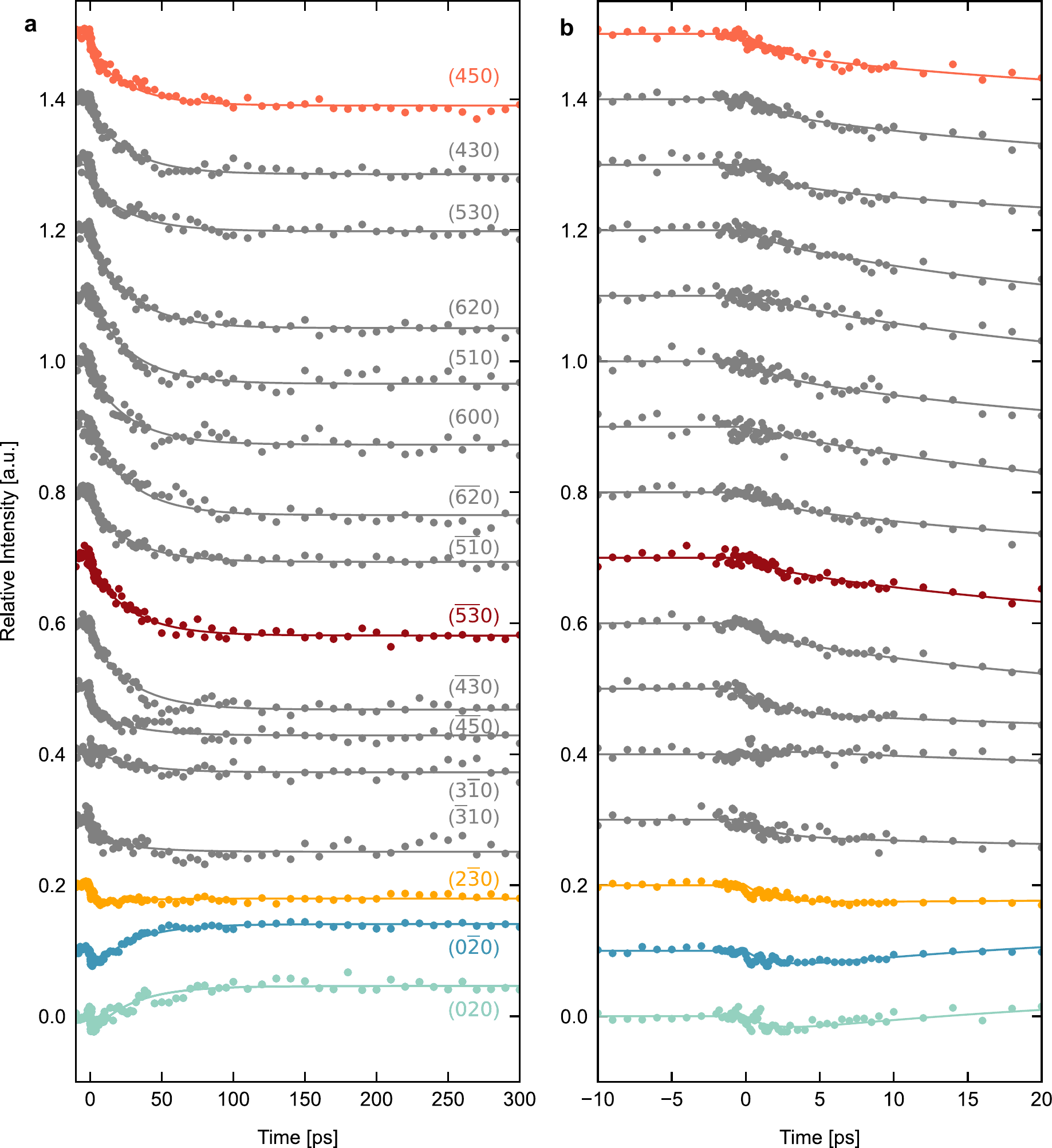}
    \caption{\textbf{Global fit to the time-resolved relative Bragg intensities.} The time constants were shared parameters for all traces, while the amplitudes were kept separate. \textbf{a} Overview of the raw data (filled circles) and fits (plain lines). \textbf{b} Zoom into the first 20 ps of the dynamics.}
    \label{fig:s1}
\end{figure}
\newpage

\newpage
\section{Supplementary Figure 4}
\begin{figure}[ht!]
    \renewcommand{\figurename}{Supplementary Figure}
    \centering
    \includegraphics[scale = 1]{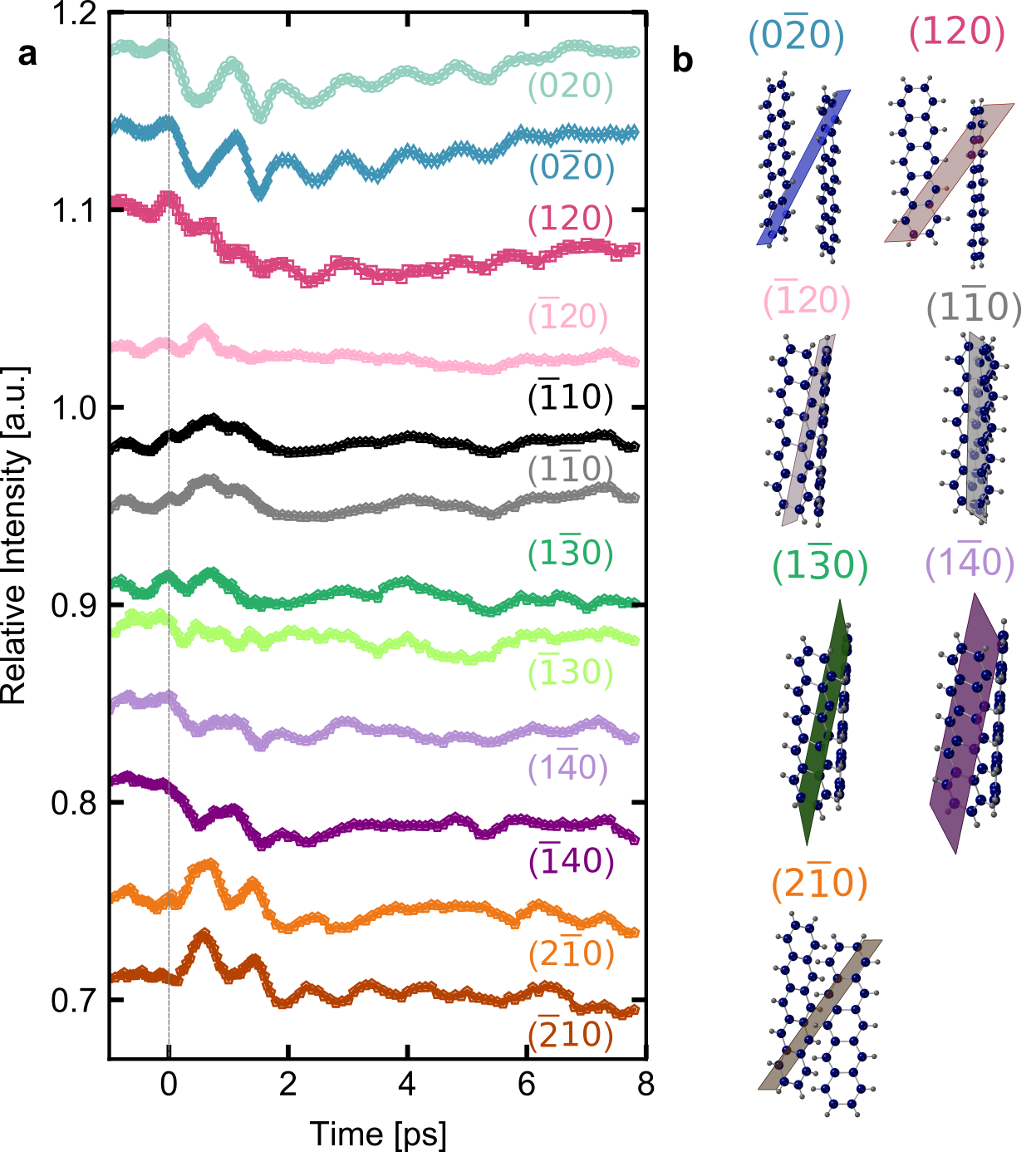}
    \caption{\textbf{Presence and absence of 1 THz oscillations in various Bragg orders.} \textbf{a} Relative intensity changes of Bragg orders in the first 8 ps. A running average with a five-point window was applied to highlight the oscillations. \textbf{b} Relative orientation of the lattice planes corresponding to the Bragg orders shown in panel \textbf{a} and the two inequivalent pentacene molecules in the unit cell. A larger coherent signal is observed in Bragg orders with corresponding diffraction planes tilted with respect to the long axis of pentacene molecules, such as (020), (0$\overline{2}$0), and (120). In contrast, weak oscillations (or absence thereof) are observed for Bragg orders with corresponding diffraction planes parallel to pentacene's long axis, such as ($\overline{1}$20) or ($\overline{1}$10).}
    \label{fig:s1}
\end{figure}

\newpage
\section{Supplementary Figure 5}
\begin{figure}[ht!]
    \renewcommand{\figurename}{Supplementary Figure}
    \centering
    \includegraphics[scale = 1.1]{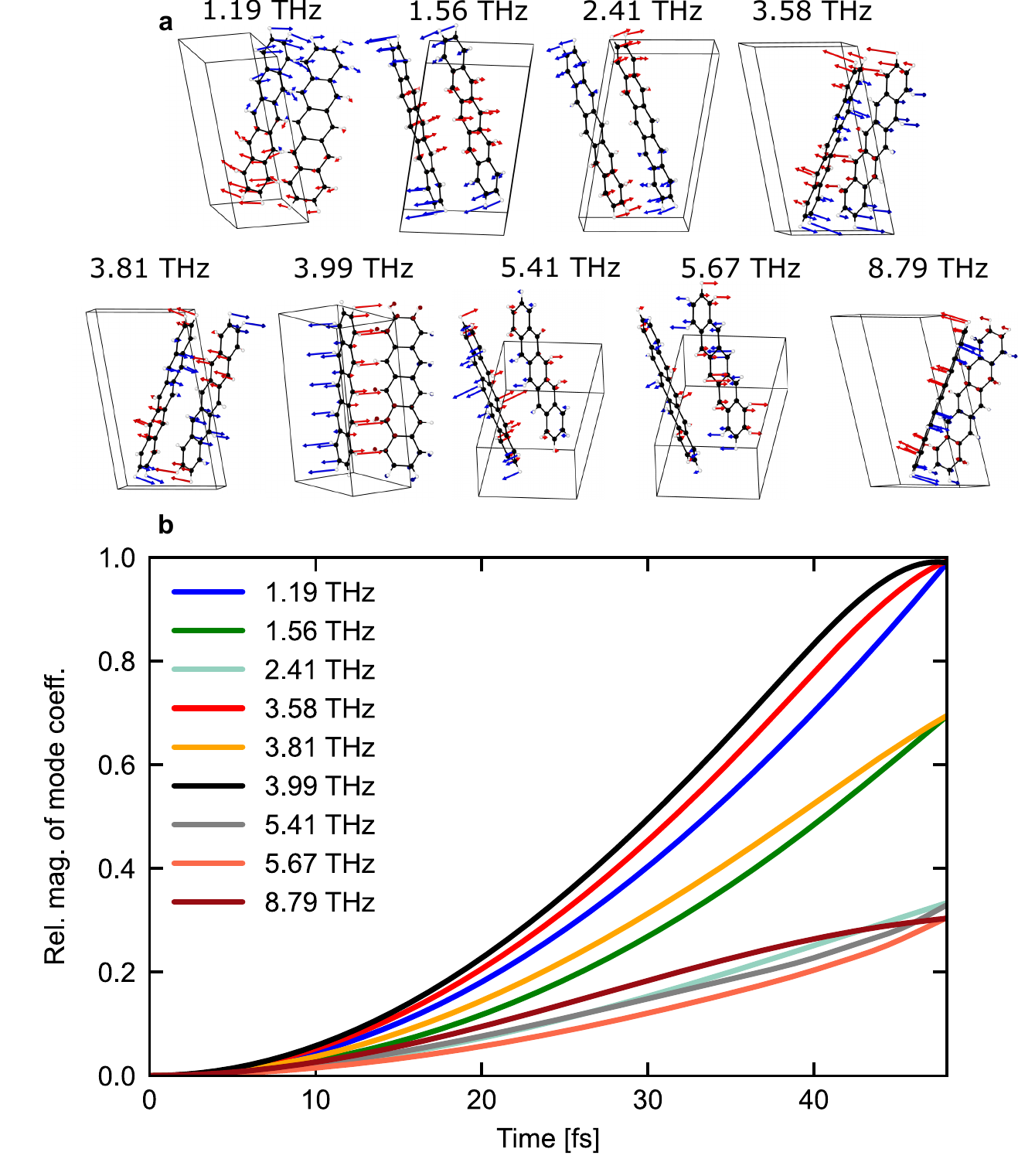}
    \caption{\textbf{Overview of the modes activated by the laser excitation in the RT-TDDFT simulations.} \textbf{a} Real-space visualization of the modes. Arrows represent the respective normal mode displacements. \textbf{b} Time evolution of the activation magnitude for the activated modes shown in panel \textbf{a}.}
    \label{fig:s7}
\end{figure}
The phonon modes shown above were obtained with the Phonopy \cite{phonopy} package using a finite atomic displacement distance of $0.002$ \AA. Only phonons at the $\Gamma$ point of the unit cell were considered. This was performed in connection with the FHI-aims program package~\cite{fhiaims}, 
where \textit{light} settings were employed for all atomic species and the Perdew-Burke-Ernzerhof exchange-correlation functional \cite{PhysRevLett.77.3865} including dispersion interactions through the  Tkatchenko-Scheffler scheme \cite{PhysRevLett.102.073005} was used (PBE+vdW).
The Brillouin zone was sampled using $5 \times 5 \times 5$ k-point mesh. The RMSD between the optimized PBE+vdW structure and the PBE+D3 optimized structure with the settings employed in the Octopus code was of 0.055 \AA. We are confident that the normal modes, if they could be calculated from Octopus, would not differ in any relevant way.

\newpage
\section{Supplementary Figure 6}
\begin{figure}[ht!]
    \renewcommand{\figurename}{Supplementary Figure}
    \centering
    \includegraphics[scale = 0.2]{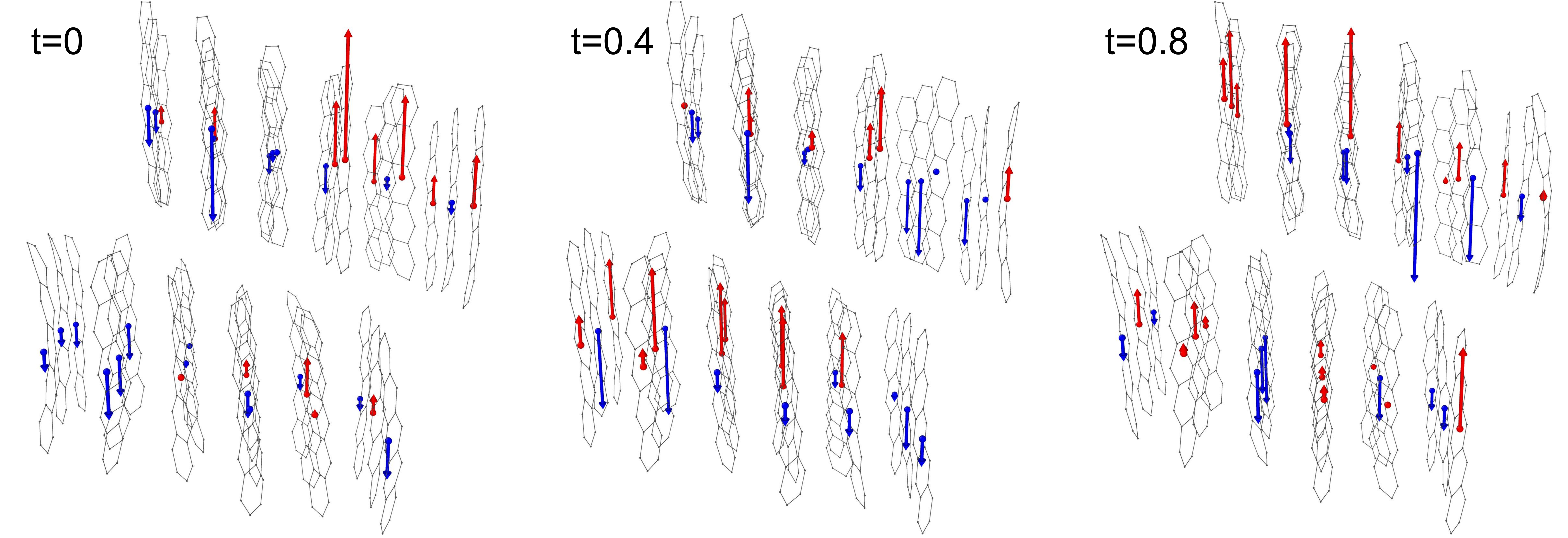}
    \caption{\textbf{Three exemplary MD snapshots.} Each snapshot corresponds to a particular time in the MD simulation and shows the whole simulation cell (3 x 6 x 2 molecules). The velocity vectors of molecular CM along $Z$ are indicated as the red and blue arrows.}
    \label{fig:s7}
\end{figure}

\newpage
\section{Supplementary Figure 7}
\begin{figure}[ht!]
    \renewcommand{\figurename}{Supplementary Figure}
    \centering
    \includegraphics[scale = 0.4]{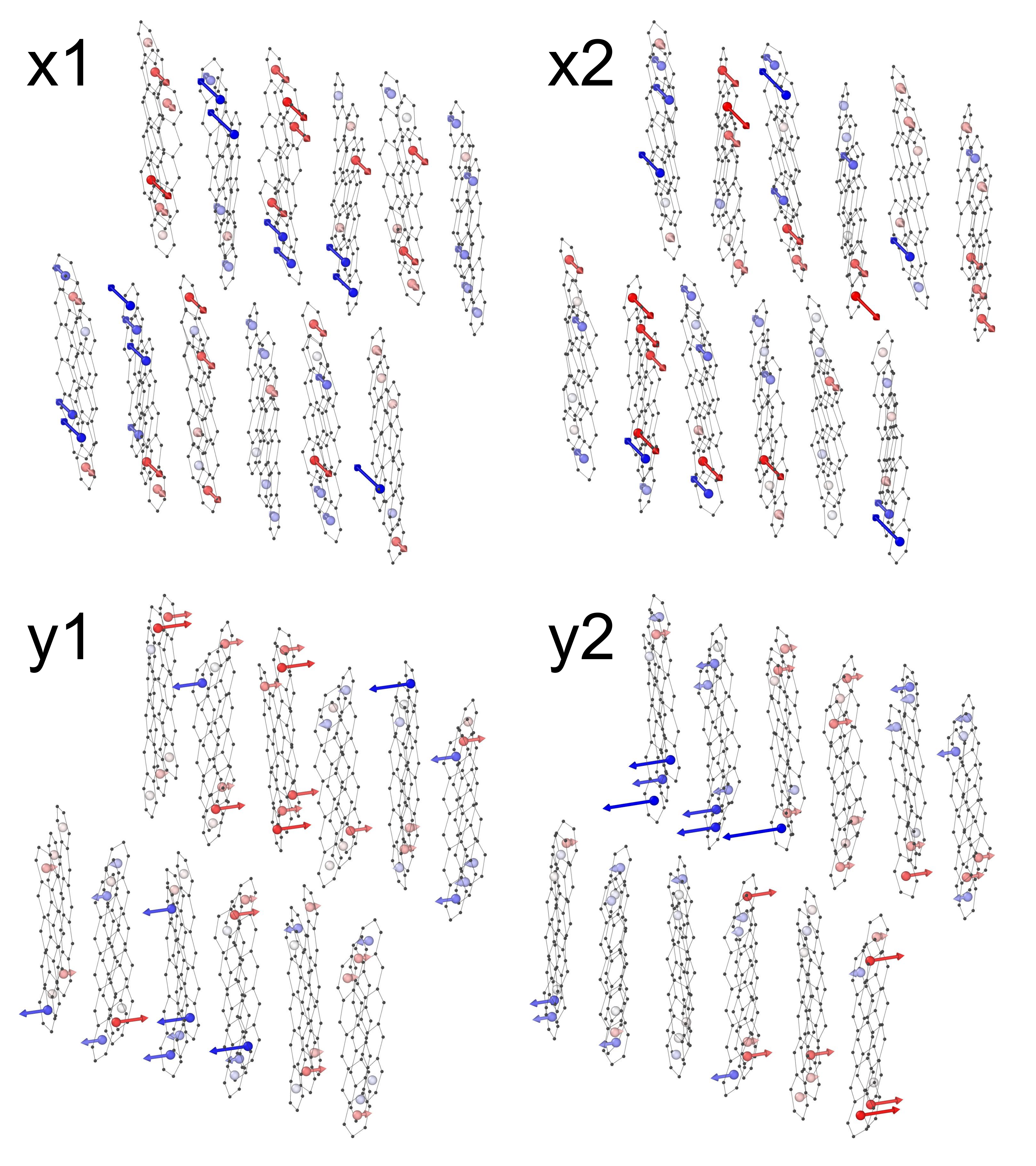}
    \caption{\textbf{Four exemplary MD snapshots.} This Figure is similar to Supplementary Figure \ref{fig:s7}, but this time the displacements of the top and botom benzene rings of each molecules along $X$ or $Y$ are indicated with red and blue arrows.}
    \label{fig:s8}
\end{figure}

\newpage

\section{Supplementary Figure 8}

\begin{figure}[ht!]
    \renewcommand{\figurename}{Supplementary Figure}
    \centering
    \includegraphics{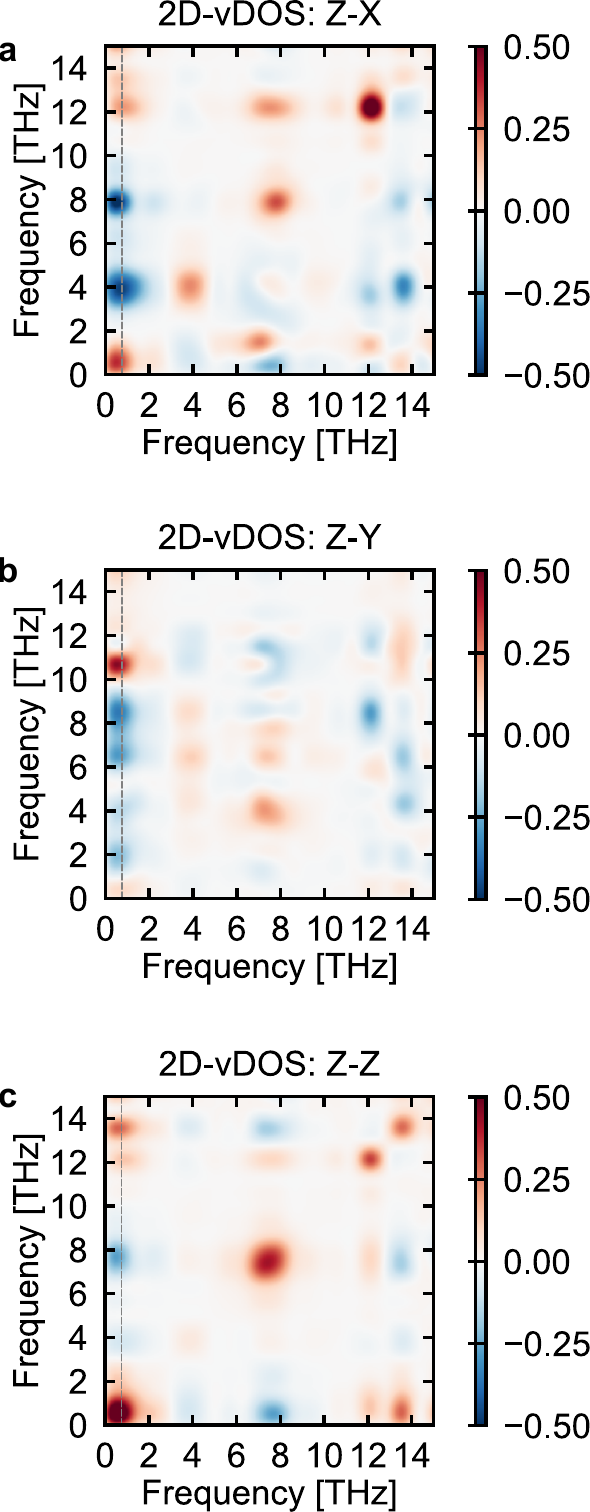}
    \caption{\textbf{2D-vDOS correlation plots extracted from MDS directly reveal phonon-phonon couplings.} \textbf{a} 2D-vDOS correlation plot obtained based on the atomic motion projected on the $Z$ and $X$ axes defined in Figure 3(f) of the main text. The grey dashed line indicates the cut shown in Figure 3(h) of the main text. \textbf{b} Same as in \textbf{a}, but for $Z$ and $Y$. \textbf{c} Same as in \textbf{a}, but for $Z$ and $Z$ (auto-correlation).}
    \label{fig:s8}
\end{figure}
\newpage

\section{Supplementary Figure 9}

\begin{figure}[ht!]
    \renewcommand{\figurename}{Supplementary Figure}
    \centering
    \includegraphics{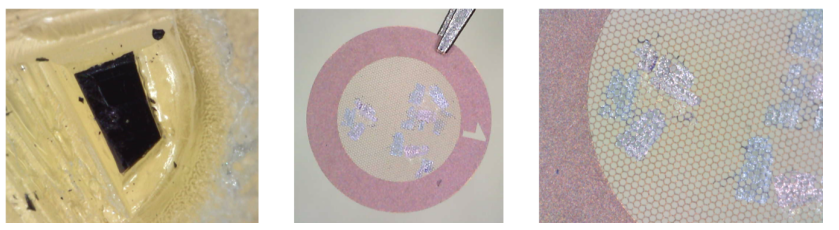}
    \caption{\textbf{a} Picture of a trimmed pentacene single crystal grown by sublimation (size of crystal: $\simeq$ 1.5 mm $\times$ 0.5 mm) \textbf{b} Microtomed freestanding sample deposited on TEM grid, employed for the FED experiments. TEM grid diameter: 3 mm. \textbf{c} Zoom of panel \textbf{b}.}
    \label{fig:s1}
\end{figure}
\newpage

\newpage
\bibliographystyle{naturemag}
\bibliography{bibliography}